\documentclass[twocolumn,aps,pra,
  showpacs,groupedaddress,amsmath,amssymb]{revtex4}

\usepackage{graphicx}% Include figure files
\usepackage{graphics}% Include figure files
\usepackage{dcolumn}% Align table columns on decimal point
\usepackage{bm}% bold math
\usepackage{color}
\usepackage{soul}

\usepackage{soul}
\usepackage[active]{srcltx}

\newcommand{\Hs}{\mathcal{H}_{\rm s}}

\begin{document}

\title{
Bringing entanglement to the high temperature limit
}

\author{Fernando Galve}
\affiliation{IFISC (CSIC - UIB), Instituto de F\'{\i}sica Interdisciplinar y Sistemas 
Complejos, Campus Universitat Illes Balears, E-07122 Palma de Mallorca, 
Spain
}
\author{Leonardo A.~Pach\'on}
\affiliation{Departamento de F\'{\i}sica, Universidad Nacional de Colombia, Bogot\'a D.C., Colombia.}
\author{David Zueco}
\affiliation{Instituto de Ciencia de Materiales de Arag\'on y
  Departamento de F\'{\i}sica de la Materia Condensada,
  CSIC-Universidad de Zaragoza, E-50012 Zaragoza, Spain.}

\date{\today}

\begin{abstract}
We show the existence of an entangled nonequilibrium state at very
high temperatures when two linearly coupled harmonic oscillators are
parametrically driven and dissipate into two independent heat
baths. This result has a twofold meaning: first, it fundamentally
shifts the classical-quantum border to temperatures as high as our
experimental ability allows us, and second, it can help increase by at
least one order of magnitude the temperature at which current
experimental setups are operated.
\end{abstract}

\date{\today}

\pacs{03.65.Yz, 03.67.Bg}

\maketitle


\textit{Introduction.}---
 Since the establishment of quantum theory in last
century there has been a long evolution on our concept of what is
quantum and to what extent it is required to explain observations in
nature. At the very beginning the reduction postulate was proposed, clearly separating between quantum
microscopic entities and classical macroscopic measuring apparatuses.
Since then macroscopic quantum phenomena such as superconductivity and coherent superposition
in Bose--Einstein condensates \cite{andrews97}, together with interference fringes of very massive molecules \cite{arndt09} have been observed. Recently a proposal to create superpositions of dielectric bodies, such as viruses up to micron size,
 inside a high finesse optical cavity has been given \cite{oriol09}. Hence the border between the classical and quantum worlds seems to be more diffuse
and intriguing than we could have conceived one century ago.

Neither the usual transition criterion of  $\hbar/S_{ch} \rightarrow
0$ (with $S_{ch}$ the characteristic action of the system) is to be
trusted, since this limit could be not completely continuous and
strong deviations have been reported in the semiclassical regime \cite{leonardo2009}.
In the more realistic situation when the system interacts with the surrounding environment, dissipation restricts purely quantum  phenomena to within the very low temperatures limit \cite{zurek03},
\begin{equation}
\label{quantumlimit}
k_B T/\hbar \omega \ll 1
\;
,
\end{equation}
where $\hbar \omega$
denotes the typical energy scale of the system and $k_B T$ the thermal
energy. Above this limit, quantum correlations are inaccessible behind a 'mask' of thermal fluctuations.

% This restriction for the manifestation of quantum phenomena, even at moderate temperatures, implies a need for a very delicate
% pre--cooling process of the system as proposed for kilogram-scale oscillators [cite] and biological entities [cite]. Here, we show that parametric driving promises a fresh and new alternative to have quantum effects even at room temperatures. 
As a consequence, observing quantum phenomena  implies the need for a
very delicate pre-cooling process.  But, {\it is there any alternative
  to cooling for being quantum?} 
In the present Letter, we defy the above classicality criterion and report the existence of a nonequilibrium entangled steady state for coupled harmonic oscillators at high temperatures, obtained through parametric driving. This result is quite fundamental, meaning that we might expect entanglement in hot highly nonequilibrium situations, as pointed out \cite{Briegel} for biological systems. Further, it could lighten the burden on quantum experiments requiring delicate pre-cooling setups.  We note that though quantum coherence can play a role in biological processes at ambient temperature \cite{scholes}, demonstration of entanglement would be a much more extreme phenomenon.
% 
% 
% This result could be of high impact to experimental requirements for manipulation of quantum systems where cooling setups are needed to reach the quantum regime, as well as expanding our notion of how fundamental and all-pervading is quantum theory. It even might be possible that biological systems\cite{Briegel} take profit from such nonequilibrium states as the one studied here.

%%%%%%%%%%%%%%%%%%%%%%%%%%%%%%%%%%%%%%%%%%%%%%%%%%%
\textit{The model and its solution.}---
%%%%%%%%%%%%%%%%%%%%%%%%%%%%%%%%%%%%%%%%%%%%%%%%%%%
In order to render our
arguments more quantitative, we study the entanglement between two
interacting identical harmonic oscillators.  Though an idealization,
it encompasses a reasonable description of a wide variety of objects
in nature such as nanomechanical oscillators \cite{NEMS}, optical \cite{aspel} and microwave
cavities \cite{Wallraff2004a}, and movable mirrors \cite{mirror} to cite some, through which we
expect to give a character of universality to the concepts that we
expose here. The Hamiltonian of the system, $\Hs$ reads
\begin{equation}
 \label{system}
\Hs = \sum_{\alpha=1} ^2 \left ( \frac{P_\alpha^2}{2 m}  +
\frac{1}{2} m \omega^2 Q_\alpha^2 \right )
+
c(t) \; Q_1 Q_2 \; ,
\end{equation}
with $m$ the mass of the oscillator, $\omega$ the frequency and $c(t)$ is the coupling coefficient.
 In what follows we assume:
\begin{equation}
\label{driving}
\frac{c(t)}{m} = c_0 + c_1 \cos (\omega_{\rm d} t)
\; ,
\end{equation}
that is, we consider a time dependent interaction, which plays a fundamental role in the creation and
survival of entanglement.

\begin{figure}[h]
\includegraphics[width=9cm, angle=-0]{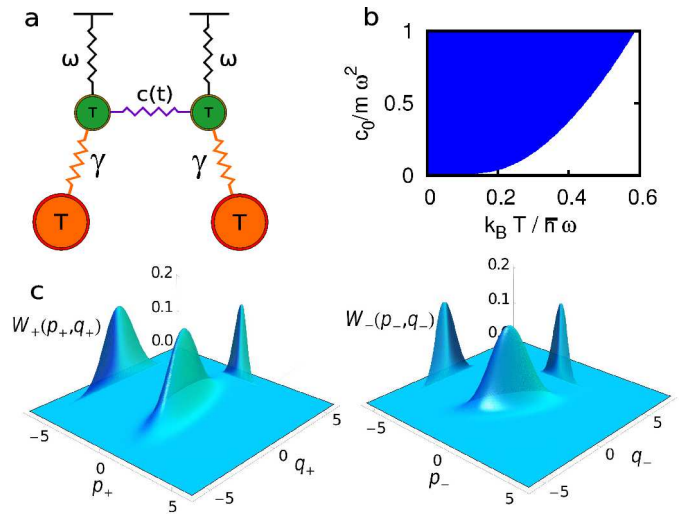}
\caption{ {\it Generation of an entangled nonequilibrium state with
    dissipative environments.} {\bf a}, The system is formed by two
  linearly coupled oscillators, initially thermalised due to each of them
 being dissipatively coupled to an environment at temperature $T$.
 Driving sinusoidally the coupling
  leads to production of entanglement even at very high
  temperatures. {\bf b}, {\it Entanglement phase diagram} for the case
  without driving. The state thermalizes to a state with no
  entanglement unless the temperature is below the quantum limit
  $k_{\rm B}T < \hbar \omega$. {\bf c}, Wigner
  phase-space representation of the normal modes. They are squeezed along
orthogonal directions, so the oscillators are entangled ($E_N\simeq0.33$). The parameters are $k_BT/\hbar\omega=10$, $\gamma=0.01\omega$, $c_1=0.5m\omega^2$, while the snapshot has been taken at time $\omega t=6$.}
\label{fig1}
\end{figure}

In any realistic scenario the system is not completely isolated from
the outside.
The most rigorous way to include dissipation is by means
of the system-bath model
\cite{Weiss1993}.  We couple the oscillators to two independent baths
(see figure \ref{fig1}a),
\begin{eqnarray}
  \label{TotalHamiltonian}
  {\cal H} &=& \Hs + \sum_{\alpha,k=1}^{2,\infty} \frac{{p_{\alpha,k}}^2 }{2{m}_k}
  + \frac{{m}_k {\omega}_k^2}{2}  \left( {x_{\alpha,k}} - \frac{{c}_k
    Q_\alpha}{{m}_k {\omega}_k^2}\right)^2 \; , \; \;
\end{eqnarray}
where the baths are modeled by an infinite
collection of harmonic oscillators \cite{Ullersma1966,
  Caldeira1983}.
%  $m_k$ and $\omega_k$ denote the mass and frequency of
% the $k$-th oscillator, which is characterized by the momentum
% $p_{\alpha,k}$ and position $q_{\alpha,k}$ coordinates
% ($[q_{\alpha,k},p_{\alpha',k'}] = {\rm i}\hbar \delta_{\alpha,\alpha'}
% \delta_{k,k'}$).  
This independence accounts for a
bath having a characteristic correlation length which exceeds the
distance between oscillators.  The opposite case
  corresponds to a model with common bath \cite{paz08} and leads to
  conservation of quantum entanglement at higher than
  $\hbar\omega/k_B$ temperatures.  This can be shown to be spurious
  since having different oscillator frequencies or bath couplings
  (and no driving) leads again to the diagram in figure \ref{fig1}b.
 Therefore we place ourselves in the most pessimistic
situation for studying entanglement \cite{chinos07}. 

The evolution for the density matrix of the two oscillators, ${\hat
  \rho}_s$, can be cast as,
\begin{equation}
  \label{EvolutionRho}
\rho_s(X_f,t) = \int {\rm d}^4 X_i J(X_f,t;X_i,0) \rho_s(X_i,0),
\end{equation}
with $X=\{Q_{1,+},Q_{1,-},Q_{2,+},Q_{2,-}\}$ and $J(X_f,t;X_i,0)$ being the
influence functional which is given in terms of a path integral expression
after tracing out the environmental degrees of freedom
\cite{Feynman1963}.
%  which propagates the system's density matrix (see
% \ref{App:DecoTotSystNModes} and \ref{App:ProFunctDenMat} for details). In
% general $J(X_f,t;X_i,0)$ is given in terms of a path integral expression
% obtained after tracing out the environmental degrees of freedom
% \cite{Feynman1963}. 
Usually, the analytical evaluation of $J(X_f,t;X_i,0)$, even for time independent
systems, is only possible in very few cases \cite{Caldeira1983,Ingold1988}.
%  However,
% in the Appendices \ref{App:DecoTotSystNModes} and \ref{App:ProFunctDenMat}
% we show that  in this case it is possible
Here,  
we have been able to derive an exact analytic expression for $J(X_f,t;X_i,0)$ in
terms of the odd and even solutions of the Mathieu oscillator (See
Appendices \ref {App:DecoTotSystNModes}  and \ref{App:ProFunctDenMat}). 
The environmental influence enters via the
spectral density $I(\omega)=\sum_j c_j^2/(2 m_j \omega_j)
\delta(\omega-\omega_j)$.
Here, we assume for simplicity {\it Ohmic} noise $I(\omega) = m \gamma
\omega$. It produces white noise in the classical limit \cite{Ingold1988}.
With our analytical result we can study the central system in any regime: low or high
temperature, strong or weak damping, deeply quantum or semiclassical
energy scales, etc., thus avoiding any
spurious approximative corrections or limitations. As a result, {\it any} system that can be
considered as two harmonic oscillators with linear coupling can be
ascribed {\it exactly} to our description.

% 
% allows us to go beyond any approximative approach, avoiding in this way any
% spurious corrections or limitations. In particular, this also means
% that we can study the central system in any regime: low or high
% temperature, strong or weak damping, deeply quantum or semiclassical
% energy scales, etc. As a result, {\it any} system that can be
% considered as two harmonic oscillators with linear coupling can be
% ascribed {\it exactly} to our description.

%%%%%%%%%%%%%%%%%%%%%%%%%%%%%%%%%%%%%%%%%%%%%%%%%%%%%%%
\textit{Entanglement computation.}---
%%%%%%%%%%%%%%%%%%%%%%%%%%%%%%%%%%%%%%%%%%%%%%%%%%%%%%%
% The driving complicate the things.  On the other hand the solution is
% possible because the total Hamiltonian system plus bath,
% Eq. (\ref{TotalHamiltonian}), is linear.  Besides, the linearity
% ensures that any gaussian initial state evolves gaussian under
% (\ref{EvolutionRho}).  By gaussian we mean that $\rho_s = {\rm det}
% \sigma \; {\rm e}^{-X \sigma^{-1} X^{t}}$, where 
Linearity of the total Hamiltonian ensures that the state is always
Gaussian, and thus
its entanglement properties are fully characterized by the covariance matrix
$\sigma_{i,j}=\langle \xi_i \xi_j+\xi_j\xi_i\rangle/2-\langle
\xi_i\rangle\langle \xi_j\rangle$ with $\xi=(Q _1,Q_2,P_1,P_2)$. An exact measure of entanglement is known for
Gaussian states, the {\it Logarithmic Negativity} $E_N$, as explained in
Appendix \ref{App:ent}.  It is computed from the covariance matrix, which can be calculated from the propagator $J(X_f,t;X_i,0)$
(see appendix \ref{App:mean}).  In what follows we will exclusively use this measure.

%%%%%%%%%%%%%%%%%%%%%%%%%%%%%%%%%%%%%%%%%%%%%%%%%%%%%%%%%%%%
\textit{Entanglement in the time independent case.}---
%%%%%%%%%%%%%%%%%%%%%%%%%%%%%%%%%%%%%%%%%%%%%%%%%%%%%%%%%%%%
In contact with an environment, each particle is asymptotically forced
into a thermal state with a temperature equal to that of the bath it
is connected to.  This state is reached independently on the initial
condition of the oscillator, which in the case of no driving [$c_1=0$
  in (\ref{driving})] leads to the entanglement characteristics shown
in figure (\ref{fig1}b). That is, any state will, after
thermalization, fall into either the blue (entangled) part or the
white (separable) part, depending only on the ratio $c_0/m\omega^2$
and the bath's temperature\cite{chinos07} \footnote {The phase diagram depends slightly on the dissipation strenght. For
  weak dissipation, as in our case, the equilibrium phase diagram is
  mostly independent on $\gamma$ \cite{Chaos2005}. In figure \ref{fig1}b we used  $\gamma=0.005\omega$}.  The
entanglement region is restricted to the so called quantum limit
$\hbar \omega < k T$, as expected from intuition, above such a
temperature each oscillator has an independent description because the
quantum state is separable \footnote{Notwithstanding each of the
  oscillators might be still regarded as quantum up to yet higher
  temperatures, we focus on entanglement since it underlies the very
  heart of the quantum weirdness.}.

%%%%%%%%%%%%%%%%%%%%%%%%%%%%%%%%%%%%%%%%%%%%%%%%%%%%%%%%%%
\textit{Entanglement creation by driving}---
%%%%%%%%%%%%%%%%%%%%%%%%%%%%%%%%%%%%%%%%%%%%%%%%%%%%%%%%%%
We sketch here a simple idea of how to produce an entangled nonequilibrium
 state at high temperatures. It may provide a huge leap in experimental requirements
 , while in addition it definitely removes temperature from the list of possible criteria
for classicality, the latter being an important theoretical topic.
 The normal mode transformation for the oscillator Hamiltonian
 (\ref{system}) reads $\widetilde H = \sum_{\alpha= \pm}
 P_{\alpha}^2/2 m + \omega_{\pm}^2 Q_{\alpha}/2$ where $Q_{\pm} =
 (Q_{1} \pm Q_{2})/\sqrt{2}$ ($P_{\pm}= (P_{1} \pm P_{2})/\sqrt{2}$)
 and $\omega_\pm^2 = \omega \pm c(t)/m$.  In the continuous variable
 setting, it is known that the maximally entangled state -a kind of
 reference state, comparison with which provides a quantification
 scheme for entanglement- is the Einstein, Podolsky, Rosen
 wavefunction \cite{EPR}. It is just the infinite squeezing limit of
 the two-mode squeezed vacuum state, in which the indeterminacies of
 $Q_+$ and $P_-$ are under the standard quantum limit set by
 Heisenberg's principle, while $Q_-$ and $P_+$ are above it (such that
 $\Delta Q_\pm/\Delta P_\pm=\exp{(\mp 2r)}/\omega^2$, with $r$ the
 so-called squeezing parameter). The opposite situation is also
 valid. Thus generation of entanglement can be provided by squeezing
 of the normal modes, which in turn can be generated through
 parametric driving of their frequencies \cite{galve09}. Both a time
 dependence in $\omega$ or $c$ will do, however the latter is better
 because it naturally provides the correct combination of squeezing
 between $\pm$ modes.  At the same time, the environment will try to
destroy quantum coherence through equilibration to the thermal state.
 Thus we have two competing effects, whose balance will
 determine whether the steady state is entangled or not.
In figure \ref{fig1} we provide an example of normal mode squeezing in
presence of the bath above the typical quantum limit
(\ref{quantumlimit}) $k_BT/\hbar\omega=10>1$.

\begin{figure}[h]
\includegraphics[width=0.45\textwidth, angle=-90]{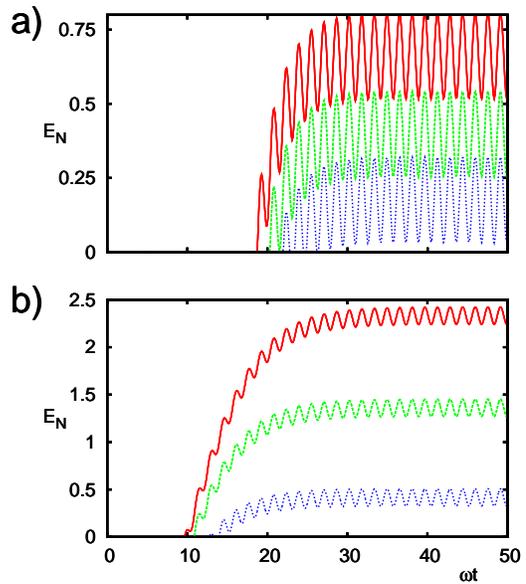}
 \caption{(Color online) {\bf a}, Time evolution of entanglement under parametric driving of the coupling at different environmental temperatures $k_BT/\hbar\omega=250$(red), $300$(green), $350$(blue), with a damping of $\gamma=0.005\omega_0$, driving amplitude of $c_1=0.5m\omega_0^2$ and driving frequency $\omega_d=2\times0.998\omega_0$. A steady entangled state is reached in a reasonable time with a significative amount of entanglement. {\bf b}, Now the temperature is kept fix , $k_BT/\hbar\omega=5$, with the same parameters, while the damping parameter is varied: $\gamma=0.005\omega_0$(red), $0.01\omega_0$(green), $0.02\omega_0$(blue).}
 \label{fig2}
 \end{figure}

In figure \ref{fig2} we summarize our results.
Indeed, we find sets of parameters where entanglement is present
at temperatures beyond the quantum limit, notice that in both figures
$k_{\rm B}T > \hbar \omega$. Starting with a thermal
  state at the bath's temperature, the system reaches after a certain
  time a nonequilibrium steady state with nonzero entanglement.
%  This
%   state is the result of a dynamical competition between the
%   decohering effect of the baths and the entanglement generating
%   contribution of the parametric coupling. 
We have chosen rather conservative
couplings to the baths, as we will explain later, and still very high
temperatures, $k_{\rm B} T \gg \hbar \omega $, can be reached.

% \begin{figure}[h]
% \includegraphics[width=7cm]{fig1b}
%  \caption{Same as figure \ref{fig1a} with a damping of $0.0005\omega_0$ and a driving amplitude of $0.8m\omega_0^2$. The temperatures here are $T=250$(black,solid), $300$(blue,dotted), $350$(red,dashed), again in units of $\hbar\omega_0/k_B$. Notice that for oscillators in the range $\omega_0=21GHz$ this would mean entanglement at room temperature.}
%  \label{fig1b}
%  \end{figure}

\begin{figure}[h]
\includegraphics[width=6.cm, angle=-90]{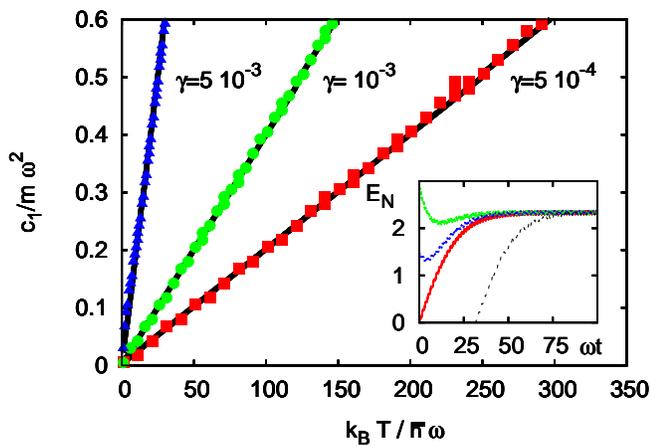}%{new_phase_diagram}%{fig2}
 \caption{'Phase diagram' of entanglement in the presence of
   parametric driving. We compare the condition (\ref{newCond})
   [lines] with the exact time evolution [dots] for different bath
   couplings $\gamma=0.005\omega$ (blue triangles), $0.001\omega$
   (green circles ) and $0.0005\omega$ (red squares). {\it Inset:} time evolution for different initial conditions, namely a two mode squeezed vaccuum state (dotted curves) with squeezing parameter $r=0$ (red), $0.5$ (blue), $1$ (green), as compared to that of an initial thermal state (black). They all converge after some tens of periods. The parameters here are $\gamma=0.001\omega$, $c_1=0.2m\omega^2$, $\omega_d=2\times0.9998\omega$ and $k_BT/\hbar\omega=10$.}
 \label{fig3}
 \end{figure}

%%%%%%%%%%%%%%%%%%%%%%%%%%%%%%%%%%%%%%%%%%%%%%%%%%%%%%%%%%%%%%%%
%  \textit{Initial conditions independence and non equilibrium steady
%    entanglement}---
%%%%%%%%%%%%%%%%%%%%%%%%%%%%%%%%%%%%%%%%%%%%%%%%%%%%%%%%%%%%%%%%
It is a remarkable fact that while the system is
 forced into a highly nonequilibrium state, a steady state of
 entanglement is reached which is independent on the initial state of
 the system. To show this effect we plot in figure \ref{fig3} (see inset) the time
 evolution of entanglement when the system starts with a two mode
 squeezed state and squeezing parameters $r=0,0.5,1$, and compare it
 to the case of an initial thermal state with the same temperature as
 the bath.

%%%%%%%%%%%%%%%%%%%%%%%%%%%%%%%%%%%%%%%%%%%%%%%%%%
 \textit {New 'phase diagram' for entanglement}---
%%%%%%%%%%%%%%%%%%%%%%%%%%%%%%%%%%%%%%%%%%%%%%%%%%
% The
%  independence of the long time entanglement dynamics with respect to
%  initial conditions makes its
%  presence at a given temperature  dependent on a few
%  parameters, 
Parametric driving yields a new asymptotic behaviour which defines a new 'phase diagram', now
dependent on four parameters: driving amplitude and frequency, temperature and the coupling to the bath.
The driving frequency is overall chosen to be $\omega_d=2\times0.998\omega$, and we also set $c_0=0$. While the optimal squeezing generation is obtained with a $\omega_d$ dependent on $\omega$ and $c_1$, the latter number seems to produce results nearly as good for different parameters, so it will be used unless otherwise stated.
%  to the bath. Therefore we can recharacterize the 'phase diagram' in
%  figure \ref{fig1} in the presence of driving.
 In figure \ref{fig3}
 we see the points which delimit the border between
 presence(left)/absence(right) of entanglement, which is linear in temperature and driving
 amplitude and, as expected, the more isolated and driven the system
 is (low $\gamma$ and high $c_1$), the higher the temperature can
 be reached. In addition to the exact result, we have plotted a simple
 estimation of the border which we explain next.

We already mentioned that the entanglement production in this system can
be viewed as a competition between the squeezing due to the driving and {\it
  mixing} because of the environment.
The rate of squeezing can be obtained from the solutions to the nondissipative driven problem. 
They have the Mathieu form $x(t)=\exp{(i\mu_M t)}\phi(t)$, where $\phi(t)$ is a
periodic function.If the Mathieu
characteristic exponent $\mu_M$ is real, they are stable, otherwise they are divergent which implies production
of squeezing at a rate $|$Im$(\mu_M)|$ (for every damped solution
there is a divergent one) \cite{Zerbe1995}.
% otherwise. Precisely the divergent character of those solutions is the
% feature which is called squeezing when translated to the quantum
% realm. 
% Therefore, the degree at which the solutions diverge is
% $|$Im$(\mu_M)|$ (for every damped solution there is a divergent one,
% so the sign does not matter). 
The rate of decoherence can be estimated from the diffusion
coefficient $D$\cite{zurek03} (see Appendix \ref{App:QME}), yielding $\gamma D\sim\gamma \ k_BT/\hbar\omega$ whenever $k_BT >\hbar\omega$. Thus by comparison of both rates we obtain the new condition under which entanglement is present:
\begin{equation}
\label{newCond}
\frac{k_BT}{\hbar\omega}\leq \frac{|\rm{Im}(\mu_M)|}{\gamma} \; ,
\end{equation}
which is seen to be a rather impressive match to the exact evolution.
The condition above should be compared with the {\it standard}
condition (\ref{quantumlimit}).  In a nutshell the driving brings in a
new quantum limit.
% 
% It must be stressed here too that the coincidence between these rough
% estimates and the actual 'quantum border' is rather impressive.

%%%%%%%%%%%%%%%%%%%%%%%%%%%%%%%%%%%%%%%%%%%%%%%%%%%%%%
\textit{Some examples}---
%%%%%%%%%%%%%%%%%%%%%%%%%%%%%%%%%%%%%%%%%%%%%%%%%%%%%%
We give next some actual examples of experiments which could profit from our strategy. However an additional comment is in order: the fact that squeezing grows approximately as $|\text{Im}(\mu_M)|t$ also means that the energy and delocalization in space are increasing exponentially in time. 
Thus checking consistency with experimental size and energy considerations is a must.

Take for example two Calcium ions, each confined in its own planar Penning traps \cite{Stahl2005}. A trap can be fabricated by nanolithography with a size of $d\sim 0.12\mu m$. If a voltage of $V=10V$ is applied, the motional frequency is $\omega\simeq 21$GHz and thus we can interpret figure \ref{fig1} as the temperature in Kelvin. A wire mediated capacitive coupling between traps allows to reduce the effective distance between ions and makes the coupling increase up to a reasonable level $c(t)=c_0=0.047m\omega_0^2$. If the frequencies are driven instead of the coupling (i.e. $\omega(t)=\omega_0+\omega_1\sin\omega_dt$), and assuming $\gamma=0.0005\omega$, we still manage to get entanglement up to $\sim50K$, while the delocalization of the oscillators is yet below the trap size, ensuring no confinement leakage. To reach room temperature a very strong coupling would be required indeed, but our method allows the experimentalist to avoid building a sub-4K (liquid Helium) setup. We believe this to be a huge experimental step.

Another example is microwave superconducting cavities
\cite{Mariantoni2008a}.  The coupling between two cavities can be
modulated placing a superconducting qubit between them. The effective
hamiltonian governing the dynamics is (\ref{system}). The typical
frequencies in these resonators are in the GigaHerz regime, operating
usually in the milikelvin range. The decoherence in these systems is $\gamma \cong 10^{-4}\omega$, or even less. However the
coupling is weak, around $10$MHz. In this case, due to the weak
coupling, the parametric driving would enhance the amount entanglement
that could be measured by nowadays
technology \cite{Menzel2010}.

Current experiments with nanomechanical resonators have these typical parameters: $\omega = 2\pi \nu = 2 \pi \times 15{\rm MHz}$,
$m = 10^{-17}{\rm kg}$, $c_1 \sim 10^{-1}m\omega^2$, and a quality factor $Q\sim 20000$, which yields a damping
$\gamma=5\times 10^{-5}\omega$ \cite{WMC08}. An entangled state can be observed at $2K$. If the
frequency can be increased a factor 10, it might reach the entangled regime in presence of liquid Helium.

In addition it is notable that the strong coupling
regime has been reached between a massive mechanical microresonator
and light \cite{aspel}.
%  Indeed, in a few more steps, two such
% resonators could in principle be coupled through light in a time
% dependent fashion.
Furthermore, a proposal for parametrically driving
the coupling between a nanomechanical resonator and a superconducting
electrical resonator has been given in \cite{simmonds08}. Thus we might well foresee that these advances could be used to measure entanglement in yet unsuspected temperature regimes in the near future, while eliminating the need for complex and costly setups to cool objects to the quantum regime.
% 
% There is right now a rush to obtain quantum behaviour with mechanical
% resonators through cooling to the ground state. We believe that our
% proposal will reduce the experimental requirements to observe the weirdest
% manifestation of quantum behavior, entanglement, at higher temperatures, which could lead to
% less costly experiments, since dilution refrigerators and expensive
% cooling setups are typically needed.

%%%%%%%%%%%%%%%%%%%%%%%%%%%%%%%%%%%%%%%%%%%%%%%%%%%%%%
\textit{Acknowledgments}---
%%%%%%%%%%%%%%%%%%%%%%%%%%%%%%%%%%%%%%%%%%%%%%%%%%%%%%
We aknowledge Peter H\"anggi and Gert-Ludwig Ingold for enlightened
discussions and advices.  We also thank the warm hospitality from the
Universit\"at Augsburg where this work was started.  DZ aknowledges
financial suport from FIS2008-01240 (MICINN), FG from COQUSYS (IFISC-CSIC), LAP from Colciencias
and the U. Nal. de Colombia.
%%%%%%%%%%%%%%%%%% bibliography %%%%%%%%%%%%%%%%%%%%%%%%%%%%%

\bibliography{quantum}

\appendix

%%%%%%%%%%%%%%%%%%%%%%%%%%%%%%%%%%%%%%%%%%%%%%%%%%%%%%%%%%%%%%%
%%%%%%%%%%%%%%%%%%  appendix%%%%%%%%%%%%%%%%%%%%%%%%%%%%%%%%%%%
%%%%%%%%%%%%%%%%%%%%%%%%%%%%%%%%%%%%%%%%%%%%%%%%%%%%%%%%%%%%%%%

\begin{widetext}
\newpage
\section{Entanglement quantification}
\label{App:ent}

Entanglement can be easily quantified for a bipartite system of continuous variables in a Gaussian state. The logarithmic negativity \cite{logneg02} gives a characterization of the amount of entanglement which can be distilled into singlets. In the case of Gaussian continuous variable states, only the covariance matrix is needed. The covariance matrix $\sigma$ is defined as
\begin{equation}
\sigma_{\xi_i\xi_j}=\langle\xi_i\xi_j+\xi_j\xi_i\rangle/2-\langle\xi_i\rangle\langle\xi_j\rangle
\end{equation}
with $\xi_i=Q_1,Q_2,P_1,P_2$. The logarithmic negativity is defined as
\begin{equation}
E_N=-\frac{1}{2}\sum_{i=1}^4\log_2[\text{Min}(1,2|l_i|)]
\end{equation}
where $l_i$ are the symplectic eigenvalues of the covariance matrix. They are simply the normal eigenvalues of the matrix $-i\Sigma\sigma$, with $\Sigma$ the symplectic matrix
\begin{equation}
\sigma=\left(\begin{array}{cc}
0&\textbf{1}_2\\
-\textbf{1}_2&0
\end{array}\right)
\end{equation}
and $\textbf{1}_2$ is the $2\times2$ identity matrix.

Whenever the logarithmic negativity of the system is zero, we have a separable state $\rho_s=\sum_i p_i \rho_1^{(i)}\otimes\rho_2^{(i)}$, and each oscillator can be described independently. In continuous variable systems, the amount of entanglement is unbounded from above, having as a limiting case the maximally entangled EPR wavefunction with $E_N\to\infty$.

\section{Decoupling the total system in normal modes}
\label{App:DecoTotSystNModes}
The Hamiltonian of the total system reads
 \begin{eqnarray}
 \Hs &=&  \frac{p_1^2}{2m} + \frac{1}{2}m\omega^2 q_1^2 + \frac{p_2^2}{2m} + \frac{1}{2}m\omega^2 q_2^2 + c(t)q_1 q_2,
\\
 {H}_{IB} &=&  \sum_{k=1}^N \frac{1}{2m_k}p_k^2 + \frac{1}{2} m_k \omega_k^2 \left( x_k - \frac{c_k q_1}{m_k \omega_k^2}\right)^2
                    + \sum_{k=1}^N \frac{1}{2{m'}_k}{p'}_k^2 + \frac{1}{2} {m'}_k {\omega'}_k^2 \left( {x'}_k - \frac{{c'}_k q_2}{{m'}_k {\omega'}_k^2}\right)^2,
\end{eqnarray}
where $c(t)= m c_0 + m c_1 \cos(\omega_{\rm d} t)$. Introducing the normal modes coordinates $x_+$ and $x_-$ defined by
\begin{eqnarray}
q_1 = \frac{1}{\sqrt{2}}\left(x_+ + x_-\right), \qquad && p_1 = \frac{1}{\sqrt{2}}\left(p_+ + p_-\right),
\\
q_2 = \frac{1}{\sqrt{2}}\left(x_+ - x_-\right), \qquad && p_2 = \frac{1}{\sqrt{2}}\left(p_+ - p_-\right),
\end{eqnarray}
$\Hs$ reads
 \begin{eqnarray}
 {\cal H}_S &=&  \frac{p_+^2}{2m} + \frac{1}{2}m\Omega_+^2(t) x_+^2 + \frac{p_-^2}{2m} + \frac{1}{2}m\Omega_-^2(t) x_-^2,
\end{eqnarray}
where $\Omega_{\pm}^2(t)=\omega^2 \pm c(t)/m$ and $H_{IB}$
 \begin{eqnarray}
  {H}_{IB} &=&
  \sum_{k=1}^N \frac{1}{2m_k}p_k^2 + \frac{1}{2} m_k \omega_k^2 x_k^2 - \frac{1}{\sqrt{2}} c_k x_k (x_+ + x_-) + \frac{c_k^2}{2 \sqrt{2} m_k \omega_k^2} (x_+ + x_-)^2
\nonumber\\
  &+& \sum_{k=1}^N \frac{1}{2{m'}_k}{p'}_k^2 + \frac{1}{2} {m'}_k {\omega'}_k^2 {x'}_k^2 - \frac{1}{\sqrt{2}} {c'}_k {x'}_k (x_+ - x_-) + \frac{{c'}_k^2}{2 \sqrt{2} {m'}_k {\omega'}_k^2} (x_+ - x_-)^2.
\end{eqnarray}
These coordinates introduce a cross-term $x_+ x_-$, which cancels out if
\begin{equation}
\frac{c_k^2}{m_k \omega_k^2} = \frac{{c'}_k^2}{{m'}_k {\omega'}_k^2}.
\end{equation}
This requirement does not means that the oscillators in the baths are identic but their  modes distributions. In the continuous limit, it implies that the spectral distributions characterizing the baths, $J_1(\omega)$ and $J_2(\omega)$, are the same. In this case,
 \begin{eqnarray}
  {H}_{IB} &=&
  \sum_{k=1}^N \left\{  \frac{1}{2m_k}p_k^2 + \frac{1}{2} m_k \omega_k^2 x_k^2 + \frac{1}{2{m'}_k}{p'}_k^2 + \frac{1}{2} {m'}_k {\omega'}_k^2 {x'}_k^2 + \frac{c_k^2}{\sqrt{2} m_k \omega_k^2} x_+^2 + \frac{c_k^2}{\sqrt{2} m_k \omega_k^2} x_-^2 \right.
\nonumber \\
  &-&\left. \left(\frac{1}{\sqrt{2}} {c}_k {x}_k +\frac{1}{\sqrt{2}} {c'}_k {x'}_k \right)x_+ - \left(\frac{1}{\sqrt{2}} {c}_k {x}_k - \frac{1}{\sqrt{2}} {c'}_k {x'}_k \right)x_- \right\}.
\end{eqnarray}
This expression suggests the introduction of new set of coordinates $\mathfrak{q}_k$ and $\mathfrak{Q}_k$ defined by
\begin{eqnarray}
\mathfrak{q}_k = \frac{1}{\lambda_k \sqrt{2}}\left( {c}_k {x}_k +{c'}_k {x'}_k \right), && \mathfrak{Q}_k = \frac{1}{\Lambda_k \sqrt{2}}\left({c}_k {x}_k - {c'}_k {x'}_k \right),
\end{eqnarray}
which can be inverted
\begin{eqnarray}
x_k = \frac{1}{\sqrt{2}{c}_k} \left( \lambda_k {\mathfrak{q}}_k + \Lambda_k {\mathfrak{Q}}_k \right), && {x'}_k = \frac{1}{\sqrt{2}{c'}_k} \left( \lambda_k{\mathfrak{q}}_k - \Lambda_k{\mathfrak{Q}}_k \right),
\\
p_k = \frac{1}{\sqrt{2}{c}_k} \left( \lambda_k {\mathfrak{p}}_k + \Lambda_k {\mathfrak{P}}_k \right), && {p'}_k = \frac{1}{\sqrt{2}{c'}_k} \left( \lambda_k{\mathfrak{p}}_k - \Lambda_k{\mathfrak{P}}_k \right).
\end{eqnarray}
After substituting in $H_{IB}$ and choosing $m_k c_k^2 = {m'}_k {c'}_k^2$ to eliminate a term proportional to $\mathfrak{p}_k \mathfrak{P}_k$, we have
 \begin{eqnarray}
  {H}_{IB} &=&
  \sum_{k=1}^N \left\{  \frac{\lambda_k^2}{2 m_k c_k^2}\mathfrak{p}_k^2 + \frac{m_k \omega_k^2 \lambda_k^2}{2c_k^2}  \mathfrak{q}_k^2 - \lambda_k  \mathfrak{q}_k x_+ + \frac{\Lambda_k^2}{2 m_k c_k^2}\mathfrak{P}_k^2 + \frac{m_k {\omega }_k^2 \Lambda_k^2}{2c_k^2}  {\mathfrak{Q}}_k^2  - \Lambda_k  \mathfrak{Q}_k x_- \right\}.
\end{eqnarray}
To obtain a more standard version of the Hamiltonian, we could redefine $m_k \rightarrow \lambda_k^2/c_k^2 \mathfrak{m}_k$ and $\omega_k^2 = c_k^2 \varpi_k^2/\lambda_k^4$ and impose $\lambda_k^2 = \Lambda_k^2$ or just by choosing $\lambda_k^2 = c_k^2 = \Lambda_k ^2$, so
 \begin{eqnarray}
  {H}_{IB} &=&
  \sum_{k=1}^N \left\{  \frac{1}{2 m_k}\mathfrak{p}_k^2 + \frac{m_k \omega_k^2}{2}  \mathfrak{q}_k^2 \pm c_k  \mathfrak{q}_k x_+ + \frac{1}{2 m_k }\mathfrak{P}_k^2 + \frac{m_k {\omega }_k^2 }{2}  {\mathfrak{Q}}_k^2  \pm c_k  \mathfrak{Q}_k x_- \right\}.
\end{eqnarray}
It means that we can conserve a small arbitrariness in the phase of the coupling by choosing different signs by $\lambda_k$ and $\Lambda_k$. However, for convenience we choose `$+$' for both.

In summary, we have
 \begin{eqnarray}
  {\cal H} &=& \frac{p_+^2}{2m} + \frac{1}{2}m\Omega_+^2 x_+^2 + \frac{p_-^2}{2m} + \frac{1}{2}m\Omega_-^2 x_-^2
  \nonumber \\
  &+& \sum_{k=1}^N \left\{  \frac{1}{2 m_k}\mathfrak{p}_k^2 + \frac{m_k \omega_k^2}{2}  \mathfrak{q}_k^2 - c_k  \mathfrak{q}_k x_+ + \frac{1}{2 m_k }\mathfrak{P}_k^2 + \frac{m_k {\omega }_k^2 }{2}  {\mathfrak{Q}}_k^2  - c_k  \mathfrak{Q}_k x_- \right\},
\end{eqnarray}
or
 \begin{eqnarray}
  {\cal H} &=& \frac{p_+^2}{2m} + \frac{1}{2}m\Omega_+^2 x_+^2 + \sum_{k=1}^N \left\{  \frac{1}{2 m_k}\mathfrak{p}_k^2 + \frac{m_k \omega_k^2}{2}  \mathfrak{q}_k^2 - c_k  \mathfrak{q}_k x_+ \right\}
  \nonumber \\
  &+&\frac{p_-^2}{2m} + \frac{1}{2}m\Omega_-^2 x_-^2
  + \sum_{k=1}^N \left\{ \frac{1}{2 m_k }\mathfrak{P}_k^2 + \frac{m_k {\omega }_k^2 }{2}  {\mathfrak{Q}}_k^2  - c_k  \mathfrak{Q}_k x_- \right\}.
\end{eqnarray}

It is quite trivial, but we have derived an effective microscopic description of our initial assumption: normal modes coupled to identic but independent baths. It worths to be mentioned that not only the baths have the same modes, $\frac{c_k^2}{m_k \omega_k^2} = \frac{{c'}_k^2}{{m'}_k {\omega'}_k^2}$, but also the coupling between the system and the bath is the same, $\lambda_k = + c_k = \Lambda_k$.

In order to complete our program an important point is left, if we want that the propagating function factorize, $J[x_+,x_-,x_+',x_-']=J[x_+,x_+'] J[x_-,x_-']$, obtaining  that each normal mode evolves actually in an independent way, we have to verify that the product by pairs of the equilibrium density matrix of the baths modes remains uncorrelated in the new coordinates.  In the current case, the transformation of coordinates reads
\begin{eqnarray}
x_k = \frac{1}{\sqrt{2}} \left( {\mathfrak{q}}_k + {\mathfrak{Q}}_k \right), && {x'}_k = \frac{1}{\sqrt{2}} \left( {\mathfrak{q}}_k - {\mathfrak{Q}}_k \right),
\\
p_k = \frac{1}{\sqrt{2}} \left( {\mathfrak{p}}_k + {\mathfrak{P}}_k \right), && {p'}_k = \frac{1}{\sqrt{2}} \left( {\mathfrak{p}}_k - {\mathfrak{P}}_k \right),
\end{eqnarray}
then, the product of the equilibrium density matrix of the $k$-th mode of each bath reads
\begin{eqnarray}
&&
\frac{1}{{\cal Z}^{k}}\left( \frac{m_k \omega_k}{2 \pi \hbar \sinh(\omega_k \hbar \beta)}\right)^{\frac{1}{2}}\exp\left[ -\frac{m_k \omega_k}{2 \pi \hbar \sinh(\omega_k \hbar \beta)} ((x_{i,k}^2 + x_{i',k}^2)\cosh(\omega_k \hbar \beta) - 2 x_{i,k} x_{i',k})\right]
\nonumber \\ &\times&
\frac{1}{{{\cal Z}'}^k}\left( \frac{{m'}_k {\omega'}_k}{2 \pi \hbar \sinh({\omega'}_k \hbar \beta)}\right)^{\frac{1}{2}}\exp\left[ -\frac{{m'}_k {\omega'}_k}{2 \pi \hbar \sinh({\omega'}_k \hbar \beta)} (({x'}_{{i},k}^2 + {x'}_{{i'},k}^2)\cosh({\omega'}_k \hbar \beta) - 2 {x'}_{i,k} {x'}_{i',k})\right]
\nonumber \\ &\rightarrow&
\frac{1}{{\cal Z}^{k}}\left( \frac{m_k \omega_k}{2 \pi \hbar \sinh(\omega_k \hbar \beta)}\right)^{\frac{1}{2}}\exp\left[ -\frac{m_k \omega_k}{2 \pi \hbar \sinh(\omega_k \hbar \beta)} (({\mathfrak q}_{i,k}^2 + {\mathfrak q}_{i',k}^2)\cosh(\omega_k \hbar \beta) - 2 {\mathfrak q}_{i,k} {\mathfrak q}_{i',k})\right]
\nonumber \\ &\times&
\frac{1}{{{\cal Z}}^k}\left( \frac{{m}_k {\omega}_k}{2 \pi \hbar \sinh({\omega}_k \hbar \beta)}\right)^{\frac{1}{2}}\exp\left[ -\frac{{m}_k {\omega}_k}{2 \pi \hbar \sinh({\omega}_k \hbar \beta)} (({\mathfrak Q}_{{i},k}^2 + {\mathfrak Q}_{{i'},k}^2)\cosh({\omega}_k \hbar \beta) - 2 {\mathfrak Q}_{i,k} {\mathfrak Q}_{i',k})\right].
\end{eqnarray}
To obtain this desired result, we had to impose $m_k = m_k'$ and $\omega_k = \omega_k'$. So, it reduces our baths to be equal in detail, we mean, oscillator by oscillator. Only at this point we can affirm that the normal modes will evolve independently. This result for the bath modes can be interpret in geometrical terms as follows: the ispotential lines of two uncoupled identic harmonic are defined by circumferences, so they are invariant under any rotation, which imply that the dynamical quantities obey exactly the same motion equations. It is important to mention that the normal modes are coupled to the bath in different coordinates than the real modes, however the introduction of the normal modes for the bath leaves the Jacobian of the transforation equals to 1, so after the trace the will generate completely equivalent results.

\section{Propagating function for the density matrix}
\label{App:ProFunctDenMat}
In normal modes, the evolution of the density matrix is governed by,
\begin{eqnarray}
\rho(x_{+,f},y_{+,f},x_{-,f},y_{-,f},t) &=& \int {\rm d}x_{+,i} {\rm d}y_{+,i} \int {\rm d}x_{-,i} {\rm d}y_{-,i} J(x_{+,f},y_{+,f},x_{-,f},y_{-,f},t|x_{+,i},y_{+,i},x_{-,i},y_{-,i},0)
\nonumber \\
&\times& \rho(x_{+,i},y_{+,i},x_{-,i},y_{-,i},t),
\end{eqnarray}
where $J(x_{+,f},y_{+,f},x_{-,f},y_{-,f},t|x_{+,i},y_{+,i},x_{-,i},y_{-,i},0)$ is the propagator of the reduced density matrix,
\begin{eqnarray}
J(x_{+,f},y_{+,f},x_{-,f},y_{-,f},t|x_{+,i},y_{+,i},x_{-,i},y_{-,i},0) &=& \int {\cal D} x_+ \int {\cal D}y_+ \int {\cal D}x_- \int {\cal D}y_-
\nonumber \\
&& \hspace{-2cm} \exp\left\{\frac{{\rm i}}{\hbar}S[x_+,x_-] - S[y_+,y_-] \right\}{\cal F}[x_+,y_+,x_-,y_-],
\end{eqnarray}
where $S[x_+,x_-]$ is the classical action and ${\cal F}[x_+,y_+,x_-,y_-]$ the influence functional. ${\cal D} x$ denotes an infinite product of measures in configuration space and implies a path integration over the paths $x_+(t)$,  $y_+(t)$, $x_-(t)$ and $y_-(t)$ with endpoints $x_+(0)=x_{+,i}$, $y(0)=y_{+,i}$, $x_-(0)=x_{-,i}$, $y_-(0)=y_{-,i}$, $x_+(t)=x_{+,f}$, $y(t)=y_{+,f}$, $x_-(t)=x_{-,f}$ and $y_-(t)=y_{-,f}$. However, at this point we have decoupled our system and we are describing it by two different harmonic oscillators coupled to identical but independent baths. So,
\begin{eqnarray}
\rho(x_{+,f},y_{+,f},x_{-,f},y_{-,f},t) &=& \int {\rm d}x_{+,i} {\rm d}y_{+,i} {\rm d}x_{-,i} {\rm d}y_{-,i} J_+(x_{+,f},y_{+,f},t|x_{+,i},y_{+,i},0) J_-(x_{-,f},y_{-,f},t|x_{-,i},y_{-,i},0)
\nonumber \\
&\times& \rho(x_{+,i},y_{+,i},x_{-,i},y_{-,i},0),
\end{eqnarray}
with
\begin{eqnarray}
J_{\pm}(x_{{\pm},f},y_{{\pm},f},t|x_{{\pm},i},y_{{\pm},i},0)  &=& \int {\cal D} x_{\pm} \int {\cal D}y_{\pm} \exp\left\{\frac{{\rm i}}{\hbar}(S_{\pm}[x_{\pm}] - S_{\pm}[y_{\pm}])\right\}{\cal F}[x_{\pm},y_{\pm}].
\end{eqnarray}
For the case of a bath modeled by harmonic oscillators \cite{Ullersma1966}, the general result for ${\cal F}[x_+,y_+]$ was derived by Caldeira and Leggett \cite{Caldeira1983} and it reads
\begin{small}
\begin{eqnarray}
\hspace{-1cm}{\cal F}[x_+,y_+]&=&\exp\left\{ - \frac{{\rm i}}{\hbar}\frac{m}{2}\left[(x_{+,i} + y_{+,i}) \int_0^t {\rm d}s \gamma(s)[x_+(s) - y_+(s)] +\int_0^t {\rm d}s \int_0^s {\rm d} u \gamma(s-u) [\dot{x}_+(u) + \dot{y}_+(u)] [x_+(s) - y_+(s)]\right]\right\}
\nonumber \\
&\times& \exp\left\{ - \frac{1}{\hbar} \int_0^t {\rm d}s \int_0^s {\rm d}u [x_+(u) - y_+(u)] K(u - s) [x_+(s) - y_+(s)]\right\},
\end{eqnarray}
\end{small}
similar expressions stands for the   ${\cal F}[x_-,y_-]$ mode, $K(s)$ denotes the noise kernel
\begin{equation}
K(s) = \int_0^{\infty} \frac{{\rm d} \omega}{\omega} \coth\left(\frac{\omega \hbar}{2 k_B T} \right) \cos(\omega s) I(\omega),
\end{equation}
wherein $k_B$ denotes the Boltzmann constant and $T$ the temperature of the bath. The friction kernel $\gamma(s)$ in terms of the spectral density reads
\begin{equation}
\gamma(s) = \frac{2}{m} \int_0^{\infty} \frac{{\rm d}\omega}{\pi} \frac{I(\omega)}{\omega} \cos(\omega s), \qquad {\rm in\, Ohmic\, case\,\,} \gamma(s) = 2 \gamma \delta(s).
\end{equation}
An identical expression stands for ${\cal F}[x_-,y_-]$. Since path integrals in $J$ are quadratic, they can be done exactly to yield
\begin{eqnarray}
J  &=& \frac{1}{N_+(t)N_-(t)}\exp\left\{\frac{{\rm i}}{\hbar}(S_+[x_+^{cl}] - S_+[y_+^{cl}] + S_-[x_-^{cl}] - S_-[y_-^{cl}]) \right\}{\cal F}[x_+^{cl},y_+^{cl}] {\cal F}[x_-^{cl},y_-^{cl}],
\end{eqnarray}
being  $N_{\pm}$ a normalization factor determined by the normalization of the propagator. To simplify further expressions, let's us to introduce the center of mass and difference variables, i.e.,
\begin{equation}
q_{\pm} = x_{\pm} - y_{\pm},\qquad Q_{\pm} =\frac{1}{2}(x_{\pm} - y_{\pm}),
\end{equation}
satisfying
\begin{eqnarray}
\ddot{q}_{\pm}(s) - \gamma \dot{q}_{\pm}(s) + \Omega_{\pm}^2(s;\varphi) q_{\pm}(s) = -2 q_{f,\pm} \gamma \delta(t-s),
 \\ && \nonumber \\
\ddot{Q}_{\pm}(s) + \gamma \dot{Q}_{\pm}(s) + \Omega_{\pm}^2(s;\varphi) Q_{\pm}(s) = -2 Q_{i,\pm} \gamma \delta(s).
\end{eqnarray}
It is important to mention that solution to these equations will be valid only for $s > 0$ and it reads \cite{Zerbe1995}
\begin{eqnarray}
q_{\pm}(s) &=& v_{1,\pm}(t,s;\varphi) q_{i,\pm} + v_{2,\pm}(t,s;\varphi) q_{f,\pm},
\\ && \nonumber \\
Q_{\pm}(s) &=& u_{1,\pm}(t,s;\varphi) Q_{i,\pm} + u_{2,\pm}(t,s;\varphi) Q_{f,\pm}.
\end{eqnarray}
% or
% \begin{eqnarray}
% q_{\pm}(s) &=& {\rm e}^{\gamma t/2} \frac{G_{\pm}(t,s)}{G_{\pm}(t,0)} q_{i,\pm} + \frac{G_{\pm}(s,0)}{G_{\pm}(t,0)} q_{f,\pm},
% \qquad
% Q_{\pm}(s) = {\rm e}^{-\gamma t/2} \frac{G_{\pm}(t,s)}{G_{\pm}(t,0)} Q_{i,\pm} + \frac{G_{\pm}(s,0)}{G_{\pm}(t,0)} q_{f,\pm},
% \end{eqnarray}
Since baths are defined by the same spectral density, then note that $\gamma$ is the same for $\pm$ cases. So we have that
\begin{eqnarray}
&& J(x_{+,f},y_{+,f},x_{-,f},y_{-,f},t|x_{+,i},y_{+,i},x_{-,i},y_{-,i},0) = \frac{1}{N(t)}
\nonumber \\
&\times& \exp\left[-\frac{{\rm i}}{\hbar}m \{ [b_{3,+}(t;\varphi) q_{+,i} - b_{4,+}(t;\varphi)q_{+,f}]Q_{+,f} + [b_{1,+}(t;\varphi)q_{+,i} - b_{2,+}(t;\varphi)q_{+,f}]\} \right]
\nonumber \\
&\times& \exp\left[-\frac{{\rm i}}{\hbar}m \{ [b_{3,-}(t;\varphi) q_{-,i} - b_{4,-}(t;\varphi)q_{-,f}]Q_{+,f} + [b_{1,-}(t;\varphi)q_{-,i} - b_{2,-}(t;\varphi)q_{-,f}]\} \right]
\nonumber \\
&\times& \exp\left[-\frac{{\rm 1}}{\hbar} \{ a_{11,+}(t;\varphi) q_{+,i}^2 +[a_{12,+}(t;\varphi) + a_{21,+}(t;\varphi)]q_{+,i}q_{+,f} + a_{22,+}(t;\varphi)q_{+,f}^2 \} \right]
\nonumber \\
&\times& \exp\left[-\frac{{\rm 1}}{\hbar} \{ a_{11,-}(t;\varphi) q_{-,i}^2 +[a_{12,-}(t;\varphi) + a_{21,-}(t;\varphi)]q_{-,i}q_{-,f} + a_{22,-}(t;\varphi)q_{-,f}^2 \} \right],
\end{eqnarray}
where $N(t)=N_+(t)N_-(t)$,
\begin{equation}
a_{ij,\pm} = \frac{1}{2}\int_0^t {\rm d}s_1 \int_0^t {\rm d}s_2 v_{i,\pm}(t,s_1;\varphi) v_{j,\pm}(t,s_2;\varphi) K(s_1 - s_2),
\end{equation}
and
\begin{eqnarray}
b_{1,\pm}(t;\varphi) &=& \dot{u}_{1,\pm} (t,0;\varphi) + \gamma, \qquad b_{2,\pm} = \dot{u}_{1,\pm} (t,t;\varphi),
\\
b_{3,\pm}(t;\varphi) &=& \dot{u}_{2,\pm} (t,0;\varphi), \qquad b_{4,\pm} = \dot{u}_{2,\pm} (t,t;\varphi),
\end{eqnarray}
Using last definitions we can express $N_{\pm}$ as $N_{\pm} = \frac{2\pi\hbar}{b_{3\pm}(t)}$. Next step is the derivation of the master equation. We based our calculation on the paper of Zerbe and H\"{a}ngii \cite{Zerbe1995} where the authors derived the exact quantum master equation for a single driven harmonic oscillator.

\section{Quantum Master Equation (QME)}
\label{App:QME}
Quantum master equation for the normal modes of the initial system reads
\begin{eqnarray}
\hspace{-1cm}{\rm i}\hbar \frac{\partial}{\partial t}\rho(x_+,y_+,x_-,y_-) &=& \left[ -\frac{\hbar^2}{2 m} \left( \frac{\partial^2}{\partial x_+^2} -  \frac{\partial^2}{\partial y_+^2} \right) + \frac{m}{2}\Omega_+^2(t;\varphi) (x_+^2 - y_+^2)\right]\rho(x_+,y_+,x_-,y_-)
\nonumber \\
&+& \left[ -\frac{\hbar^2}{2 m} \left( \frac{\partial^2}{\partial x_-^2} -  \frac{\partial^2}{\partial y_-^2} \right) + \frac{m}{2}\Omega_-^2(t;\varphi) (x_-^2 - y_-^2)\right]\rho(x_+,y_+,x_-,y_-)
\nonumber \\
&-&\frac{{\rm i}\hbar \gamma}{2}(x_+ - y_+)\left( \frac{\partial}{\partial x_+} -  \frac{\partial}{\partial y_+} \right) \rho(x_+,y_+,x_-,y_-)
+ {\rm i} D_{+,pp}(t,0)(x_+^2 - y_+^2)\rho(x_+,y_+,x_-,y_-)
\nonumber \\
&-&\frac{{\rm i}\hbar \gamma}{2}(x_- - y_-)\left( \frac{\partial}{\partial x_-} -  \frac{\partial}{\partial y_-} \right) \rho(x_+,y_+,x_-,y_-)
+ {\rm i} D_{-,pp}(t,0)(x_-^2 - y_-^2)\rho(x_+,y_+,x_-,y_-)
\nonumber \\
&-&\frac{\hbar}{m}[D_{+,xp}(t,0) + D_{+,px}](x_+ - y_+)\left( \frac{\partial}{\partial x_+} +  \frac{\partial}{\partial y_+} \right) \rho(x_+,y_+,x_-,y_-)
\nonumber \\
&-&\frac{\hbar}{m}[D_{-,xp}(t,0) + D_{-,px}](x_- - y_-)\left( \frac{\partial}{\partial x_-} +  \frac{\partial}{\partial y_-} \right) \rho(x_+,y_+,x_-,y_-),
\end{eqnarray}
where
\begin{eqnarray}
D_{\pm,pp}(t,0) = 2\left(b_{4,\pm} + \frac{\dot{b}_{2,\pm}}{b_{2,\pm}}\right)a_{22,\pm} - \dot{a}_{22,\pm} + 2\frac{\dot{b}_{2,\pm}b_{4,\pm}}{b_{2,\pm}b_{3,\pm}} - \frac{b_{4,\pm}}{b_{3,\pm}} \dot{a}_{12,\pm},
\\
D_{\pm,px}(t,0) = D_{\pm,xp}(t,0) = -\frac{1}{b_{3,\pm}}\dot{a}_{12,\pm} + a_{22,\pm} + \frac{\dot{b}_{2,\pm}}{b_{2,\pm}b_{3,\pm}}a_{12,\pm}.
\end{eqnarray}

For small values of $\hbar$, $D_{\pm,px}(t,0)$ and $D_{\pm,pp}(t,0)$ can be written as \cite{dillen09}
\begin{align}
D_{\pm,pp}(t,0) &= \frac{m \gamma}{\beta} + \frac{2m^2\gamma \Lambda}{\beta}\left(\Omega_{\pm}^2(t) - \gamma^2\right),
\\
D_{\pm,px}(t,0) &= \frac{2m\gamma^2 \Lambda}{\beta},
\end{align}
where $\Lambda = \hbar^2 \beta^2 / 24 m$.

%For the undriven case at very low dissipative rates $\gamma$
%\begin{equation}
%D_{\pm,pp} = \gamma \langle p_{\pm}^2 \rangle = \frac{m^2 \hbar \Omega_{0,\pm}}{2 m}
%\left(1+\frac{2}{e^{\beta \hbar \Omega_{0,\pm} } - 1}\right)   \left(1 + \frac{\beta  \gamma  \hbar}{2 \pi ^2} \frac{  \Im\left(\psi ^{(1)}\left(\frac{i \beta
%  \Omega_{0,\pm} \hbar }{2 \pi }\right)\right) }{{\rm coth} \left(\frac{\beta  \Omega_{0,\pm} \hbar }{2 \pi }\right)}\right),
%\end{equation}
%where $\psi(z)$ is the polygamma function and $\psi^{(1)}(z)$ denotes its first derivative. $\Omega_{0,\pm} $ stands for the frequency at zero driving. At high temperature,
%\begin{equation}
%D_{\pm,pp} = \gamma \langle p_{\pm}^2 \rangle = \frac{\Omega_{0,\pm} \hbar \gamma m}{2} + m \gamma k_B T - \frac{2 m \gamma^2 k_B T}{\omega_D},
%\end{equation}
%where $\omega_D$ is the Drude cutoff frequency.

\section{Mean values and variances}
\label{App:mean}
\begin{eqnarray}
\langle f(x_{\pm}) \rangle = \int {\rm d} Q_{f,\pm} f(Q_{f,\pm}) \rho(Q_{f,\pm}, q_{f,\pm}=0,t)
\end{eqnarray}
The first moments read in terms of the initial values $\langle x_{\pm}(t_0=0) \rangle =\langle (x_{\pm,0}) \rangle$ and,
\begin{equation}
\langle x_{\pm}(t) \rangle = [f_{2,\pm}(t)-\frac{\gamma}{2}f_{1,\pm}(t) ]\langle (x_{\pm,0}) \rangle + \frac{1}{m} f_{1,\pm}(t) \langle (p_{\pm,0}) \rangle
\end{equation}
\begin{eqnarray}
\langle p_{\pm}(t) \rangle &=& m \frac{{\rm d}}{{\rm d}t} \langle x_{\pm}(t) \rangle
\\
&=& m [{\dot f}_{2,\pm}(t)-\frac{\gamma}{2}{\dot f}_{1,\pm}(t) ]\langle (x_{\pm,0}) \rangle + \frac{1}{m} {\dot f}_{1,\pm}(t) \langle (p_{\pm,0}) \rangle
\end{eqnarray}
The evolution of $\langle p_{\pm}(t) \rangle$ is discontinuous at $t_0=0$, i.e., $\lim_{t\rightarrow0^+}\langle p_{\pm}(t) \rangle = \langle p_{\pm,0} \rangle - m \gamma \langle x_{\pm,0} \rangle /2$ es in general not equal to $\langle p_{\pm,0} \rangle$. This instantaneous jump of $\langle p_{\pm}(t) \rangle$ can be removed with an environmental cutoff $\omega_c$, or a non-factorizing initial state \cite{Ingold1988}. The variances are obtained accordingly. They are given by
\begin{eqnarray}
 \sigma_{x_{\pm}x_{\pm}}(t)=\left( f_{2,\pm} -\frac{\gamma}{2}f_{1,\pm}  \right)^2\sigma^0_{x_{\pm}x_{\pm}} +  \frac{2}{m} f_{1,\pm} \left( f_{2,\pm} -\frac{\gamma}{2}f_{1,\pm} \right)\sigma^0_{x_{\pm}p_{\pm}} + \frac{1}{m^2}f_{1,\pm}^2 \sigma^0_{p_{\pm}p_{\pm}} + \frac{2\hbar}{m}f_{1,\pm}^2 a_{11,\pm},
\end{eqnarray}
\begin{eqnarray}
 \sigma_{x_{\pm}p_{\pm}}(t) &=&
m\left[ f_{2,\pm}{\dot f}_{2,\pm}
- \frac{\gamma}{2}\left(f_{1,\pm}{\dot f}_{2,\pm} + {\dot f}_{1,\pm}f_{2,\pm} - \frac{\gamma}{2} f_{1,\pm}{\dot f}_{1,\pm} \right) \right]\sigma^0_{x_{\pm}x_{\pm}}
\nonumber \\
&+&  \left( f_{1,\pm} {\dot f}_{2,\pm} + {\dot f}_{1,\pm} f_{2,\pm}  -\gamma {\dot f}_{1,\pm} f_{1,\pm} \right)\sigma^0_{x_{\pm}p_{\pm}} + \frac{1}{m^2}{\dot f}_{1,\pm} \sigma^0_{p_{\pm}p_{\pm}} + 2\hbar \left(f_{1,\pm} {\dot f}_{1,\pm} a_{11,\pm} + f_{1,\pm} a_{12,\pm}\right),
\end{eqnarray}
\begin{eqnarray}
 \sigma_{p_{\pm}p_{\pm}}(t) &=& m^2\left( {\dot f}_{2,\pm} -\frac{\gamma}{2}{\dot f}_{1,\pm}  \right)^2\sigma^0_{x_{\pm}x_{\pm}} + 2 m {\dot f}_{1,\pm} \left( {\dot f}_{2,\pm} -\frac{\gamma}{2}{\dot f}_{1,\pm} \right)\sigma^0_{x_{\pm}p_{\pm}} + {\dot f}_{1,\pm}^2 \sigma^0_{p_{\pm}p_{\pm}}
\nonumber \\
&+& 2\hbar m\left({\dot f}_{1,\pm}^2 a_{11,\pm} + 2 {\dot f}_{1,\pm} a_{12,\pm} + a_{22,\pm} \right),
\end{eqnarray}
where we omitted the arguments of the functions $ a_{ij,\pm}$ and $f_{i,\pm}$ for better lucidity. Here we note two missprints in \cite{Zerbe1995}, one is the presence of a global factor $\frac{1}{2}$ in the last term of $\sigma_{x_{\pm}p_{\pm}}$ and the other is in the last term of $\sigma_{x_{\pm}p_{\pm}}$, in \cite{Zerbe1995} it reads $2\hbar m\left(2 {\dot f}_{1,\pm}^2 a_{11,\pm} + {\dot f}_{1,\pm} a_{12,\pm} + a_{22,\pm} \right)$. Due to the discontinuity at $t=0$, variances at $t=0^+$ jump to
\begin{eqnarray}
\sigma_{x_{\pm}x_{\pm}}(t_{0^+}) &=& \sigma_{x_{\pm}x_{\pm}}^0,
\\
\sigma_{x_{\pm}p_{\pm}}(t_{0^+}) &=& -\gamma \sigma_{x_{\pm}x_{\pm}}^0 + \sigma_{x_{\pm}p_{\pm}}^0,
\\
\sigma_{p_{\pm}p_{\pm}}(t_{0^+}) &=& \gamma^2 \sigma_{x_{\pm}x_{\pm}}^0 - 2 \gamma \sigma_{x_{\pm}p_{\pm}}^0 +\sigma_{p_{\pm}p_{\pm}}^0,
\end{eqnarray}
where $t_{0^+}$ means $	\lim t\rightarrow 0^+$.
\end{widetext}

\end{document}